\documentclass[conference]{IEEEtran}
\IEEEoverridecommandlockouts
\usepackage{amsmath,amssymb,amsfonts}
\usepackage{algorithmic}
\usepackage{graphicx}
\usepackage{textcomp}
\usepackage{xcolor}
\def\BibTeX{{\rm B\kern-.05em{\sc i\kern-.025em b}\kern-.08em
    T\kern-.1667em\lower.7ex\hbox{E}\kern-.125emX}}

\usepackage[
backend=biber,
style=ieee,
sorting=none,
url=false,
doi=false,
isbn=false,
eprint=false
]{biblatex}

\usepackage{orcidlink}

\usepackage{multirow}
\usepackage[normalem]{ulem}
\useunder{\uline}{\ul}{}
\usepackage{booktabs}

\DeclareFieldFormat{journaltitle}{#1\isdot}
\DeclareFieldFormat{conferencetitle}{#1\isdot}
\DeclareFieldFormat{title}{#1\isdot}

\begin{document}

\title{Trustworthy artificial intelligence in the energy sector: Landscape analysis and evaluation framework \\
}

    \author{\IEEEauthorblockN{Sotiris Pelekis\IEEEauthorrefmark{1}\orcidlink{0000-0002-9259-9115}, Evangelos Karakolis\IEEEauthorrefmark{1}\orcidlink{0000-0003-2833-3088}, George Lampropoulos\IEEEauthorrefmark{1}\orcidlink{0009-0003-9237-2476}, \\
    Spiros Mouzakitis\IEEEauthorrefmark{1}\orcidlink{0000-0001-9616-447X}, Ourania Markaki\IEEEauthorrefmark{1}\orcidlink{0000-0001-6416-2824}, Christos Ntanos\IEEEauthorrefmark{1}\orcidlink{0000-0002-5162-6500}, and Dimitris Askounis\IEEEauthorrefmark{1}\orcidlink{0000-0002-2618-5715}} \\
\IEEEauthorblockA{\IEEEauthorrefmark{1}School of electrical and computer engineering,
National Technical University of Athens, Greece}}

\maketitle

\begin{abstract}
The present study aims to evaluate the current fuzzy landscape of Trustworthy AI (TAI) within the European Union (EU), with a specific focus on the energy sector. The analysis encompasses legal frameworks, directives, initiatives, and standards like the AI Ethics Guidelines for Trustworthy AI (EGTAI), the Assessment List for Trustworthy AI (ALTAI), the AI act, and relevant CEN-CENELEC standardization efforts, as well as EU-funded projects such as AI4EU and SHERPA. Subsequently, we introduce a new TAI application framework, called E-TAI, tailored for energy applications, including smart grid and smart building systems. This framework draws inspiration from EGTAI but is customized for AI systems in the energy domain. It is designed for stakeholders in electrical power and energy systems (EPES), including researchers, developers, and energy experts linked to transmission system operators, distribution system operators, utilities, and aggregators. These stakeholders can utilize E-TAI to develop and evaluate AI services for the energy sector with a focus on ensuring trustworthiness throughout their development and iterative assessment processes. 
\end{abstract}


\section{Introduction}
The energy sector in Europe is undergoing a notable transformation with the integration of artificial intelligence (AI) technologies \cite{Serban2020ArtificialCities}. AI is being applied in various fields including demand forecasting \cite{Pelekis2022InPerformance, Pelekis2023ADrivers, Tzortzis2023TransferSeries}, flexibility management \cite{Michalakopoulos2024APrograms}, demand response \cite{Pelekis2023ADrivers}, EV charging optimization \cite{Chen2022ApplicationReview}, load dissagregation \cite{Schirmer2023Non-IntrusiveReview}, and predictive maintenance \cite{Mahmoud2021TheReview}, presenting various implications and ethical considerations as most of them treat personal data or personally identifiable data (PII) collected from smart meters. Specifically, the integration of AI in the energy sector brings about various potential dangers that need to be addressed. These dangers entail the following: i) Cybersecurity risks: the increased use of AI in energy systems exposes vulnerabilities to cyberattacks \cite{Gunduz2019AnalysisApplications, Hao2022AdversarialGrids}, which have surged significantly in recent years. These attacks, leveraging AI, pose a threat to the security and stability of energy infrastructure \cite{Yilmaz2021AvoidingLearning} ii) Data management risks: AI applications in energy systems can lead to challenges related to data management, including data privacy, integrity, and accessibility \cite{Gharaibeh2017SmartTechnologies}. Ensuring secure and efficient data handling becomes crucial to prevent potential risks associated with AI deployment. iii) Loss of human oversight \cite{Enqvist2023HumanWhom}: overreliance on AI systems without adequate human oversight can result in errors or malfunctions that may have severe consequences in the energy sector. Maintaining a balance between automation and human control is essential to mitigate risks. iv) Environmental impact: the significant energy demand of AI technologies, particularly in data centers, raises concerns about environmental sustainability. The power consumption of AI systems, especially GPUs used for processing, contributes to increased energy consumption and water usage, impacting the overall environmental footprint of the energy sector \cite{Wu2022SustainableOpportunities}. vi) Safety concerns: faulty predictions or malfunctions in AI systems used for energy management can pose safety risks, potentially leading to accidents or failures in critical infrastructure \cite{Amodei2016ConcreteSafety}. Ensuring the reliability and accuracy of AI applications is crucial to prevent safety hazards within the energy sector. vii) Discrimination and promotion of energy inequality against justice: AI systems are prone to erroneous decisions influenced by biased or poisoned datasets, hence producing results that are unfair and promote societal inequality \cite{Noorman2023AIJustice} (e.g. demand response schemes that favor high-income consumers, against low-income ones that cannot afford smart meters).

Addressing these dangers requires a comprehensive approach that combines robust cybersecurity measures \cite{Syrmakesis2023AEstimation, Li2020SearchFromFree:Detection}, effective data management strategies, human oversight mechanisms, adaptable technology solutions, sustainable energy practices, and a focus on safety and reliability in AI deployment within the energy industry. Towards this direction, trustworthy AI (TAI) is crucial in ensuring the responsible and ethical deployment of AI systems within the energy domain. In the context of the energy sector, the European Commission (EC) emphasizes the critical importance of TAI to drive innovation while ensuring ethical and responsible AI deployment. The EC's Ethics guidelines for TAI (EGTAI) \cite{High-LevelExpertGrouponAI2019EthicsFuture}, alongside the White Paper on Artificial Intelligence \cite{EuropeanCommission2020WhiteTrust} outline key strategies to promote TAI, focusing on fundamental rights, safety, and transparency, explainability, and robustness. This commitment to ethical AI aligns with the European Union's (EU) broader efforts to establish clear guidelines and regulations for AI technologies, emphasizing the need for transparency, accountability, and human oversight in AI applications within various sectors, including energy. By prioritizing TAI principles, the European Commission aims to foster a regulatory environment that supports innovation while safeguarding ethical standards and legal compliance in the energy domain. In the same direction, the EC has issue the Artificial Intelligence ACT, which has already been decided to be adopted as an EU law, also resulting to its final draft in 2024 \cite{EuropeanCommission2024EUDraft}.

The I-NERGY project \cite{Karakolis2022ArtificialProject} is an EU-funded project focused on advancing AI in the energy sector through innovative services \cite{Pelekis2022InPerformance, Pelekis2023TargetedTechniques, Pelekis2023DeepTSF:Forecasting, Michalakopoulos2023Data-drivenNetworks, Tzortzis2023TransferSeries}. By deepening AI research, deploying decentralized governance for energy data, and providing technology enablers for advanced AI-based solutions, I-NERGY demonstrated a strong commitment to innovation and sustainability. The project's emphasis on AI interaction and TAI showcases its dedication to ensuring responsible and ethical AI deployment in the energy sector. Through its efforts to empower European AI innovation and reinforce the service layer of the AI-on-demand platform \cite{AI4EU2020AI4EUPlatform}, I-NERGY has been at the forefront of driving advancements in TAI applications for the energy industry also extending its impact beyond the energy sectors to new coming projects such as EnerShare \cite{EnerShare2022EneshareProject} and iTrust6G \cite{iTrust6G2023ITrust6GProject} that aim to put it in practice to new applications and domains respectively.

The present work, which has initially been introduced in the I-NERGY project, aims at the following: i) provide an overview of regulatory frameworks and initiatives in the context TAI; ii) identify possible implications and ethical issues of AI systems within applications of the electrical energy sector including smart power grid and smart buildings. Therefore, it can provide guidance for the instantiation of AI systems in the energy domain and facilitate the compliance with the ethical and legislative requirements. 

Ethical considerations often underlie the law and represent the common rationale, and thus overlap with the law to a certain degree. As such, they may serve as guidance where the law is not entirely adapted to new phenomena, e.g., where technology enables practices, which the legislator had not anticipated. In newly emerging fields, such as AI, legislation may indeed not cover sufficiently all ethical implications or may have not clear rules on them, thus hampering compliance. In this context, the contributions of this paper are as follows: 
\begin{itemize}
    \item a concise analysis of the research landscape and regulatory framework related to the energy domain within the EU. To the best of our knowledge no previous work has attempted such a focused analysis of TAI frameworks in the energy domain.
    \item a novel TAI framework (named E-TAI) developed in the I-NERGY project is proposed for the energy domain, introducing practices mainly inspired from EGTAI. Said framework facilitates the development and latter evaluation of TAI systems in terms of TAI compliance. Despite some attempts to introduce TAI in practice \cite{Li2023TrustworthyPractices, Otoum2021EnablingSolutions}, no previous work has proposed such a generalized methodology in the EPES domain.
\end{itemize}

\section{Landscape of Trustworthy AI in the EU}
The advancement of AI is accompanied by great opportunities for economic development and addressing societal challenges. The potential of AI is expected to radically transform the energy sector, revolutionize the way that Electric Power and Energy Systems (EPES) community is undertaking the business processes and have a significant impact on society and environment. However, AI can put pressure on ethical values and fundamental rights that drive our lives and our societies. In this context, it is necessary for the development of relevant AI systems to be in line with ethical principles and requirements, preventing any harmful implications. In the same direction, it is crucial to identify all possible ethical issues and implications within AI projects and systems, thus mitigating the associated risks and maximizing trustworthiness, impact, and sustainability. To achieve this, we examine they key initiatives and ongoing efforts around TAI as a preliminary tool before unveiling the E-TAI framework.  

\subsection{Ethics Guidelines for Trustworthy AI}
With trust being a prerequisite for human-centered AI, EC set up the High-Level Expert group on AI (AI HLEG) in June 2018 to provide advice on its Strategy \cite{High-LevelExpertGrouponAI2019EthicsFuture}. The AI HLEG prepared and published in 2019 the EGTAI, where key concepts and requirements of TAI are prescribed, and EC highlighted these requirements through its Communication on “Building Trust in Human Centric Artificial Intelligence” \cite{EuropeanCommission2019BuildingIntelligence}. In parallel to HLEG, the European AI Alliance \cite{EuropeanCommission2018TheAlliance} was established, bringing together multiple stakeholders for an open discussion on AI, including its impacts.

At first, it is meaningful to examine the term “Trustworthy AI”. According to AI HLEG, AI systems should meet the following conditions to be deemed trustworthy: i) Be in line with all applicable laws and regulations (lawful) ii) Comply with ethical principles and values. (ethical) iii) Be robust concerning both a technical and social perspective (robust). AI HLEG puts fundamental rights, enshrined in EU treaties, EU Charter for fundamental Rights \cite{EuropeanUnion2012CharterUnion} and other international human rights laws, at the center of a TAI approach. These rights contribute to the lawful dimension of AI systems as they are legally binding, but they also form the basis for the ethical principles and guidelines that AI systems should follow. AI HLEG describes the categories of fundamental rights that are suitable for AI systems. The reflection of these rights will raise awareness regarding the aspects that should be considered considered when assessing an AI system from an ethical perspective. Therefore, these rights are the following: i) respect for human dignity; ii) freedom of the individual; iii) respect for democracy, justice and the rule of law; iv) equality, non-discrimination and solidarity; v) citizens’ rights. Grounded on fundamental rights, EGTAI also lists the four principles, that AI systems, must respect in order for their development and operation to be deemed trustworthy. These principles, as described in the EGTAI document, are: i) human autonomy; ii) prevention of harm; iii) fairness; iv) explicability.

\subsubsection{Requirements for TAI}
The requirements for TAI, as defined by the EGTAI, constitute a non-exhaustive list of 7 key equally weighted factors that should be considered through the implementation stage of AI systems. These requirements derive from the previously mentioned 4 ethical principles and come to put them in practice within the implementation stages of an AI system. The 7 requirements are described in detail in the EGTAI, and are briefly summarised as follows.
\paragraph {Human agency and oversight} AI systems should uphold human autonomy, decision-making, fundamental rights, democracy, and equality. Assessing potential threats to rights before development is crucial. Providing users with information and tools for informed decisions is essential. Human oversight varies based on application and risks, requiring defined government mechanisms like human-in-the-loop, human-on-the-loop, and human-in-command.
\paragraph {Technical robustness and safety} AI systems must be technically robust to prevent harm and ensure human integrity. Resilience to attack and security is crucial to safeguard against vulnerabilities. Preventing misuse by malicious actors is essential. Fallback plans and safety measures are vital components of robust systems. Identifying risks and implementing necessary measures are key processes. Accuracy, reliability, and reproducibility are critical considerations for AI systems to function effectively under various conditions.
\paragraph {Privacy and data governance} Data governance is vital for privacy in AI systems, ensuring protection and lawful use of collected or generated information. Maintaining data integrity and quality is essential to address biases before model training. Access policies should be clearly defined to restrict data access to authorized personnel under specific conditions.
\paragraph{Transparency} AI systems must prioritize traceability, documenting datasets, algorithms, and decision-making processes for error identification and correction. Explainability is crucial, ensuring AI processes and decisions are understandable to humans. Users should be informed about interacting with AI systems and their capabilities and limitations.
\paragraph{Societal and environmental well-being} For AI systems to be trustworthy, they must promote inclusion and diversity, addressing biases in datasets and algorithms to prevent prejudice and discrimination. Oversight processes and diverse perspectives are essential to mitigate bias risks. Designing AI solutions for accessibility according to relevant standards and principles ensures equal access for all, including individuals with disabilities. Involving stakeholders enhances the development of trustworthy systems.
\paragraph{Diversity, non-discrimination and fairness} AI systems should be designed, developed and operate in the most environmentally friendly manner and measures towards this direction are encouraged. In addition, the social impact and the effects that AI systems may have on people’s mental and physical health should be considered and monitored. Attention should also be paid to the possible impacts of AI systems on society and democracy. 
\paragraph{Accountability} Establishing mechanisms for ensuring responsibility and accountability of AI systems is essential. Auditability, assessment, and evaluation reports enhance trustworthiness. Identifying, reporting, and mitigating negative impacts through impact assessments are crucial. When conflicts arise, considering appropriate trade-offs is necessary, documenting them and evaluating their ethical implications. If tradeoffs violate ethical principles, the development of such systems should be prohibited. Providing mechanisms for affected parties to seek redress is also important.

\subsubsection{Technical and non-technical methods}
To further help the realisation of Trustworthy AI, the Guidelines suggest a set of technical and non-technical methods for implementing the defined requirements. 

The technical methods for TAI, as described in the EGTAI, include: i) Architectures defining acceptable behaviors and monitoring compliance. ii) Ethics integration from design phase to safeguard data and prevent risks. iii) Emphasizing explainability for user understanding and system reliability. iv) Comprehensive testing and validation across the AI system lifecycle to ensure consistency with requirements. v) Establishing quality of service indicators for evaluating system development from various perspectives like security and usability.

Besides the technical methods, EGTAI recommend non-technical methods to enhance AI trustworthiness, including: i) Compliance with regulations and legislative frameworks. ii) Updating policies and codes of conduct based on EGTAI guidance. iii) Utilizing standards for information, rules, and quality management. iv) Certification for transparency to benefit public understanding. v) Implementing governance frameworks for accountability. vi) Promoting education and awareness for an ethical mindset among stakeholders. vii) Encouraging stakeholder participation and social dialogue for equitable access to AI benefits. viii) Engaging diverse teams in AI system development for realistic and objective outcomes.

\subsubsection{Assessment list for Trustworthy AI}
\begin{figure}[htbp]
\centering
\includegraphics[width=0.75\columnwidth, height=4.5cm]{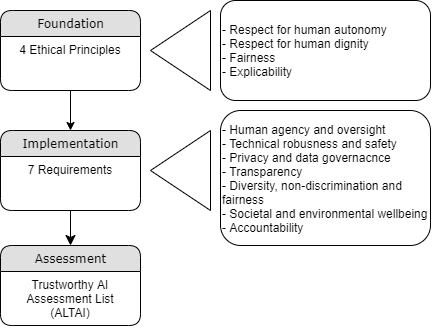}
\caption{The TAI framework as established by HLEG}
\label{fig:altai}
\end{figure}

AI HLEG presented on 17 July 2020 the final Assessment list for Trustworthy AI (ALTAI) \cite{EuropeanAIAlliance2020ALTAIPortal}. This list makes ethics central to the development of AI systems. It acts as a self-evaluation tool for assessing AI systems under the key requirements defined in the EGTAI. The list contains a set of questions relevant to the requirements that provide guidance for their practical implementation. In addition, this list raises awareness around the potential impact and risks of the proposed AI systems and the kind of measures that can be taken to mitigate these risks. The ALTAI comes to seal the process of TAI by enabling the inspection and validation of the final AI system, hence completing the lifecycle of TAI development as shown in Fig. \ref{fig:altai}.

\subsection{Artificial intelligence ACT}
On 21 April 2021, the European Commission unveiled the draft AI act, a new proposal for an EU regulatory framework on AI. The draft act signals the European Commission’s shift from a soft-law approach, as indicated by the publication of its non-binding EGTAI and Policy investment recommendations \cite{EuropeanCommission2019PolicyIntelligence} towards a legislative one. The draft AI act has been the first ever attempt to enact horizontal regulation of AI, applicable to all AI systems, developed, placed on the market or used in the Union, establishes a technology-neutral definition of the latter in EU law, and lays down for them a classification with different requirements and obligations tailored on a 'risk-based approach', whereby legal intervention is dependent upon the concrete level of risk. In particular, the draft AI act distinguishes between AI systems posing (i) unacceptable risk, (ii) high risk, (iii) limited risk, and (iv) low or minimal risk. Under this approach, AI applications would be regulated only as strictly necessary to address specific levels of risk. Note here that the final draft of the ACT has been issued on 26 January 2024 and is pending vote from the EU parliament within March 2024. The AI act aims to ensure that AI systems used within its borders are safe, transparent, traceable, non-discriminatory, and environment friendly. It emphasizes human oversight over AI systems to mitigate the risk of adverse outcomes. The act introduces specific obligations for both providers and users, tailored to the risk level associated with different AI technologies.


\subsubsection{High-risk systems}
Given that the power system is a critical infrastructure and serves vital human needs, most AI systems related to the latter can be considered high-risk according to the AI act. In this direction, the act mandates an ex-ante conformity assessment for high-risk AI systems among others. AI products and services governed by existing product safety legislation, will fall under the existing third-party conformity assessment structures and regulatory frameworks that already apply. Providers of AI systems that are not currently governed by explicit regulatory frameworks are obliged to conduct their own conformity assessment (self-assessment) and register their systems in an EU database managed by the Commission before placing them on the market or putting them into services. Proportionate obligations are also placed on users and other participants across the AI value chain.
Additional technical and auditing requirements for high-risk AI systems concern: i) risk management; ii) data and data governance; iii) technical documentation: production of detailed technical documentation, including around system architecture, algorithmic design, and model specification; iv) record-keeping; v) transparency and provision of information to users; vi)	human oversight system designed to maintain human oversight at all times and prevent or minimize risks to health and safety or fundamental rights, including an override or off-switch capability; vii) accuracy, robustness and cybersecurity.

\subsubsection{CEN-CENELEC standardization}
To assist the implementation of the AI act, in May 2023 the European Commission released a standardization request to the CEN-CENELEC for supporting the implementation of the AI act provisions \cite{CEN-CENELEC2023CEN/TCStandardisation}. The standards requested by the Commission include logging, data quality and governance, accuracy, robustness, risk management, quality management, conformity assessment, cybersecurity, human oversight, transparency. However, the current high-level, domain-agnostic requirements set at EU-level need to be customized to specific domains to ensure effective and trustworthy application. There is also an issue with the interoperability between CEN-CENELEC and ISO standards, posing a significant barrier for the interoperability of AI systems in EU and international markets. This issue could burden companies, complicating compliance with both sets of standards and hindering market interactions.

\subsection{AI4EU project}
Significant contribution to the field of ethics on AI is also provided by the AI4EU project \cite{AI4EUproject2019AI4EUAI-on-Demand}. One of the strategic objectives of AI4EU has been to promote European values for Ethical, Legal, Socio-Economic and Cultural (ELSEC) issues in AI. Societal concerns over the use and misuse of AI are addressed with the organization of an ethical, legal, socio-economical and gender-aware observatory, providing the AI community as well as European and national authorities with detailed, accurate and up to date information regarding the consequences of use and misuse of AI. While AI technologies can provide great benefit for European society, misuse can pose grave risks. In this direction, AI4EU has created the AI4EU Ethical Observatory working to assure respect of European values and to assure that respect for these values provides an important competitive advantage both within the EU and in larger international markets.
The AI4EU project also developed an abbreviated assessment framework \cite{AI4EU2021AssessmentAI}, based on the ALTAI. This abbreviated assessment list is mainly meant to assess the AI applications shared through the AI4EU catalog, but it also can support organisations perform a ‘quick scan’ of their AI systems. This list can be used as a self-assessment tool to quickly identify the relevant elements of trustworthy AI and the level of compliance to these elements. It will help determine the level of impact of the AI applications and provide options to balance different tensions and interests. The effort of AI4EU towards trustworthy AI will be now continued by its follow-up project called DeployAI.

\subsection{SHERPA project}
The Energy Trilemma \cite{WorldEnergyCouncil2024WorldIndex}, as outlined by the World Energy Council, includes Energy Security, Energy Equity, and Environmental Sustainability. It is essential to balance these factors for economic success and competitiveness. Finding solutions to the trilemma promotes technological advancement and raises ethical questions in pursuit of an ideal balance. While AI systems can help tackle the Energy Trilemma, it is important to prioritize ethical concerns. Current studies tend to prioritize functionality over ethics, underscoring the need for responsible ethical considerations. The H2020 project SHERPA investigated the ethical and human rights implications of Smart Information Systems (SIS) with a specific case study on the ethical issues of using said systems in electricity grids. Note here that SHERPA did not explicitly treat AI systems, however its contribution is still invaluable to the AI landscape in a smart grid context. Among other results, a case study \cite{Hatzakis2019SmartEth-ics} has been published, within the context of this project, analysing the principal ethical issues that occur in the use of SIS in electricity grids. These ethical issues entail: 
i) Privacy and informed consent: 
AI services utilize detailed household energy data from smart meters, raising concerns about privacy and informed consent. Newcomers, such as aggregators keep an eye on household energy consumption for programs promoting energy (e.g demand response), which could expose personal details and routines. Managing these concerns requires finding a middle ground between monitoring activities at home and respecting privacy rights, both in terms of technology and regulations. In this direction, there are important questions to address regarding the ability to opt out for individuals in low-income households or shared spaces, particularly when access to affordable energy is at stake.
ii) Energy security: 
Cyber-attacks on the energy grid, pose significant risks due to vulnerabilities created by sensors and IoT networks. These attacks can disrupt energy distribution and infrastructure, therefore impacting people's well-being and causing economic issues at a national level. ENISA \cite{ENISA2024TheCybersecurity} reported global losses of up to 1.69 billion euros in 2018 due to cyber-attacks against the energy grid \cite{NikolakopoulosTheodoros2016TheCIIs}. 
iii) Energy equity and affordability:
Smart grids aim to manage energy abundance and rising demand \cite{Sovacool2015EnergyApplications}, yet they raise ethical concerns related to energy justice. Affluent consumers may initially benefit more from AI and smart grids, potentially leading to unequal access and bias on AI model training. Dynamic pricing supported by AI could further increase energy poverty \cite{EuropeanCommission2024EnergyHub} and inequalities, contradicting the EU's goals of promoting energy equity and affordability. 
iv) Sustainability: 
AI technologies and smart devices play a crucial role in smart grids, supporting the EU's decarbonization efforts by enabling precise electricity flow management and increased renewable energy sources (RES) integration. However, the energy-intensive nature of smart grid equipment and AI models results in significant electricity consumption and CO$_2$ emissions \cite{Lacoste2019QuantifyingLearning}. Studies like \cite{Sias2017UCLACalifornia} indicate that the integration of smart devices in grids may not always lead to a reduction in CO$_2$ emissions from a life cycle assessment (LCA) perspective \cite{Reijnders2012LifeEmissions}.

\section{E-TAI -- Methodological framework for Trustworthy AI in the energy domain}
This section introduces the the proposed methodological framework for TAI assessment within AI applications the energy sector, named E-TAI. Specifically, the framework initially proposes a non-exhaustive set of guidelines for AI system stakeholders (developers, pilot users) on how to identify the ethical risks during its development lifecycle. Subsequently, the methodological framework is established so that the users and developers can monitor the developed AI solutions with respect to trustworthiness. The EGTAI and ALTAI are in the heart of the E-TAI, which however incorporates recommendations and guidelines from multiple of the aforementioned initiatives such as the aforementioned AI4EU (methodology) and SHERPA (energy domain) projects. Regarding the AI act, it is of utmost importance for the users of the framework to take into consideration that most smart grid-related AI systems should be considered as high-risk systems. Nonetheless, it has been used in a complementary manner, given its still ongoing / draft status. Finally, the General Data Protection Regulation (GDPR) also plays a central role within the proposed guidelines. 

\subsection{Supporting guidelines for the identification and management of ethical risks within AI systems}

This section is structured based on the 7 requirements proposed EGTAI. Each requirement is accompanied by a table that identifies potential risks along with proposed (non-exhaustive) technical or non-technical methods  for their mitigation. Specifically, the objective of this section is to identify potential ethical risks and implications of AI related to the energy domain. The approaches and tools proposed can be used as a starting point for the application of the E-TAI methodology. However, the reader is strongly encouraged to revisit the EGTAI and the ALTAI and only use the proposed guidelines as an auxiliary tool for the lifecycle of their AI system.

\subsubsection{Human agency and oversight}
AI systems should be aimed at supporting human agency and decision-making and are not meant to replace them. Table \ref{tab:humanagency} lists the potential technical and non-technical methods that can be used for the compliance of energy domain AI systems with this requirement.

\begin{table}[htbp]
\centering
\renewcommand{\arraystretch}{1.3} 
\begin{tabular*}{\linewidth}{p{1.35cm}|p{5.25cm}|p{0.9cm}}
\toprule
\textbf{Risks} & \textbf{Methods \& Tools} & \textbf{Method Type} \\ \hline
\midrule

\multirow{1}{\linewidth}{User misperception, deception, addiction, manipulation} 

& Fundamental rights impact assessment. Available tools: 
\newline i) AI \& Equality: Human Rights Toolbox \cite{WomenattheTable2023AIToolbox}; 
\newline ii) An approach for Fundamental Rights Impact Assessment to Automated Decision-Making \cite{Janssen2020AnDecision-making}; 
\newline iii) EU tool for fundamental rights \& human rights \cite{EuropeanJustice2023FundamentalTool}. & Non-technical \\ \cline{2-3}

~ & Consider the Article 22 of GDPR (Automated individual decision making) & Non-technical \\ \cline{2-3}

~ & Establish governance mechanisms (human-in-the-loop, human-on-the-loop, human-in-command \cite{High-LevelExpertGrouponAI2019EthicsFuture}) to ensure human oversight. & Non-technical \\ \cline{2-3}

~ & Keep the end-user informed on the degree of automation and the reasoning behind decisions. Ensure intuitive user interface & Technical \\
\bottomrule
\end{tabular*}
\caption{Risks and proposed methods with respect to human agency and oversight}
\label{tab:humanagency}
\end{table}

\subsubsection{Technical robustness and safety}
Technical robustness and safety is mostly a technical requirement that is mainly related with the resilience of an AI system to cyber attacks. Stakeholders should continuously monitor ethical issues and risks related to energy security, given that all AI solutions strongly rely on the extensive usage of SIS which will always be at a certain degree vulnerable to cyber-crime. This requirement is critical for the power grid and related services as vital infrastructures can be affected in case of malfunctioning or downtime of its components. In the same direction, all three attributes— confidentiality, integrity, availability— of information security should be ensured with respect to the data ingested and used by the AI systems (note here that this statement is strongly linked to requirement of privacy and data governance as well). In this context, Table \ref{tab:robustness} lists the risks relating to this requirement and mainly proposes some technical methods for their mitigation. 

\begin{table}[htbp]
\centering
\renewcommand{\arraystretch}{1.3} 
\begin{tabular*}{\linewidth}{p{1.1cm}|p{5.5cm}|p{0.9cm}}
\toprule
\textbf{Risks} & \textbf{Methods \& Tools} & \textbf{Method Type} \\ \hline
\midrule

\multirow{1}{\linewidth}{Cyber incidents and cyber attacks / Technical Faults} 
 & Consider ISO standards and certifications (EU cybersecurity ACT \cite{EuropeanParliament2019RegulationAct}) & Non-technical \\ \cline{2-3}

~ & Fill the Machine Learning Canvas. This can help validate that all the stages of the ML Lifecycle have been addressed ensuring technical robustness of the AI system. & Non-technical \\ \cline{2-3}

~ & Red Teaming / Penetration Testing & Technical \\ \cline{2-3}

~ & Adversarial robustness, and privacy measures (e.g. ART \cite{Nicolae2018AdversarialV1.0.0}, Pysyft \cite{OpenMined2023PySyft:Server}) & Technical \\ \cline{2-3}

~ & Identity and access management (e.g. Keycloak \cite{Camposo2021SecuringKeycloak}), AI model logging and versioning (e.g. MLflow \cite{Alla2021}) & Technical \\ \cline{2-3}

~ & Strong documentation & Technical \\ 

\bottomrule
\end{tabular*}
\caption{Risks and proposed methods with respect to technical robustness and safety}
\label{tab:robustness}
\end{table}

\subsubsection{Privacy and data governance}
The prevention of harm to privacy requires data governance procedures that ensure the confidentiality and integrity of the data to be used and processed by AI systems. In this context, Table \ref{tab:privacy} indicatively lists several potential risks and methods, associated with the requirement of privacy and data governance within ML systems in the energy domain.

\begin{table}[htbp]
\centering
\renewcommand{\arraystretch}{1.3} 
\begin{tabular*}{\linewidth}{p{1.1cm}|p{5.5cm}|p{0.9cm}}
\toprule
\textbf{Risks} & \textbf{Methods \& Tools} & \textbf{Method Type} \\ \hline
\midrule

\multirow{1}{\linewidth}{Privacy breaches, Reidentification, Bad data quality / Integrity loss} 
 & Conduct a Data Protection Impact Assessment (DPIA) – Proposed in the GDPR \cite{EuropeanParliament2016RegulationRegulation} & Non-technical \\ \cline{2-3}

~ & Fill a Data Ethics Canvas \cite{OpenDataInstitute2021TheCanvas} to gain awareness of the data processes of the AI system & Non-technical \\ \cline{2-3}

~ & Privacy-by-design (anonymisation, data minimisation, encryption, differential privacy, federated learning etc.).  & Technical \\ \cline{2-3}

~ & Apply quality controls on ingested energy datasets. (e.g automated scripts on sensors/ smart meters) & Technical \\ \cline{2-3}

~ & Consider low granularity / resolution of datasets. Consider timesteps larger than 15-30 minutes (instead of storing real time measurements). & Technical \\ \cline{2-3}

~ & Identity and access management (e.g. Keycloak \cite{Camposo2021SecuringKeycloak}) & Technical \\

\bottomrule
\end{tabular*}
\caption{Risks and proposed methods with respect to privacy and data governance}
\label{tab:privacy}
\end{table}

\subsubsection{Transparency}
Transparency refers to 3 main pillars: i) traceability ii) explainability and iii) communication regarding the limitations of the system. Table \ref{tab:transparency} gives an overview of indicative risks and mitigation actions that can be indicatively adopted by stakeholders of ML services.

\begin{table}[htbp]
\centering
\renewcommand{\arraystretch}{1.3} 
\begin{tabular*}{\linewidth}{p{1.1cm}|p{5.5cm}|p{0.9cm}}
\toprule
\textbf{Risks} & \textbf{Methods \& Tools} & \textbf{Method Type} \\ \hline
\midrule

\multirow{1}{\linewidth}{Inability to contest a decision, fake expectations, untransparent decisions, excessive trust in the AI system} 
 & Usage of methodologies to ensure transparency of data management processes and their usefulness for users (e.g Datasheets for Datasets \cite{Gebru2021DatasheetsDatasets}) & Non-technical \\ \cline{2-3}

~ & Communication of the abilities and limitations of the AI system across its various users. The user should know that she is interacting with AI & Non-technical \\ \cline{2-3}

~ & Strong documentation of the development process (models and datasets) and decision-making mechanisms to ensure traceability and ability of the user to contest decisions.  & Technical \\ \cline{2-3}

~ & Consider interpretability / explainability related software components within the AI system, such as Lime \cite{RibeiroMarco2016Lime:Classifier}, Shap \cite{SHAP2018SHAP:Model} & Technical \\ \cline{2-3}

~ & In case of deep learning systems, consider using interpretable models such as N-BEATS \cite{Oreshkin2020N-BEATS:Forecasting} and Temporal Fusion Transformer (TFT) \cite{Lim2019TemporalForecasting} & Technical \\

\bottomrule
\end{tabular*}
\caption{Risks and proposed methods with respect to transparency}
\label{tab:transparency}
\end{table}

\subsubsection{Diversity, non-discrimination and fairness}
It is of utmost importance that AI systems avoid discriminatory bias. Discriminatory bias refers to systematic errors in AI algorithms that lead to decisions against specific groups of people. The bias can be caused by (i) inappropriately trained AI algorithms, (ii) datasets that are not representative of reality (due to bad data collection or pre-processing). Table \ref{tab:fairness} lists indicative risks and methods with respect to the development of fair AI systems.

\begin{table}[htbp]
\centering
\renewcommand{\arraystretch}{1.3} 
\begin{tabular*}{\linewidth}{p{1.6cm}|p{5.0cm}|p{0.9cm}}
\toprule
\textbf{Risks} & \textbf{Methods \& Tools} & \textbf{Method Type} \\ \hline
\midrule

\multirow{1}{\linewidth}{Discrimination, deteriorations of social inequalities, marginalisation, unfair competition} 
 & Adopt participatory approaches to development (working groups, questionnaires etc.) & Non-technical \\ \cline{2-3}

~ & Monitoring of data collection process to ensure the  representativeness and quality (inclusion of different social groups) & Technical \\ \cline{2-3}

~ & Consider removing variables that introduce bias in the models & Technical \\ \cline{2-3}

~ & Comply with accessibility standards, hence allowing all groups of people to utilise the AI system & Technical \\ \cline{2-3}

~ & Consider producing a bias report through fairness metrics and open-source tools such as AIF360 \cite{Bellamy2018AIBias}, Aequitas \cite{Saleiro2018Aequitas:Toolkit}, Fairtest \cite{Tramer2017FairTest:Applications}, Themis-ML \cite{Bantilan2017Themis-ml:Mitigation} & Technical \\

\bottomrule
\end{tabular*}
\caption{Risks and proposed methods with respect to diversity, non-discrimination and fairness}
\label{tab:fairness}
\end{table}

\subsubsection{Societal and environmental well-being}
Based on a study from McKinsey \cite{ChuiMichael2019UsingProgramme}, AI systems can accelerate most of the UN Sustainable Development Goals (SDGs). Nonetheless, R. Vinuesa et al. show in \cite{Vinuesa2020TheGoals} that AI may act as an enabler on 134 targets (79\%) across all SDGs, while 59 targets (35\%, also across all SDGs) may experience a negative impact from the development of AI. In this context, AI system stakeholders should ensure the alignment of their AI systems with the UN’s sustainability goals. Amongst them energy poverty, energy efficiency and climate change are some of the most crucial and are potentially strongly linked with power grid related AI systems. Table \ref{tab:wellbeing} indicatively lists risks and proposed methods relating to social and environmental well-being.

\begin{table}[htbp]
\centering
\renewcommand{\arraystretch}{1.3} 
\begin{tabular*}{\linewidth}{p{1.15cm}|p{5.35cm}|p{0.9cm}}
\toprule
\textbf{Risks} & \textbf{Methods \& Tools} & \textbf{Method Type} \\ \hline
\midrule

\multirow{1}{\linewidth}{Hinder the realisation of UN SDG’s, deterioration of human skills (social skills, job loss), reinforcement of authoritarian behaviour, social scoring systems, acceleration of climate change
} 
 & Monitoring and alignment with UN’s SDG’s and European legislation and directives (GDPR, AI act, Data act, Cybersecurity ACT etc.) & Non-technical \\ \cline{2-3}

~ & Establish procedures for assessing the CO$_2$ emissions of ML. The procedures should encompass: i) compute related impacts that include the electricity used for AI computations alongside the embodied emissions associated with the infrastructure (top-down approach); ii) immediate impacts that are tied to the short-term outcomes of ML systems (e.g. some AI application might decrease the cost of grid emissions but lead to the increase of energy consumption overall); iii) structural or "system-level" CO$_2$ emissions induced AI applications that can have broader societal implications beyond their immediate impact. Some tools have already been proposed by Kaack et al. \cite{Kaack2022AligningMitigation} regarding: a) reports for measuring ML model energy usage and CO$_2$ emissions; b) metrics for reporting model accuracy as a function of computational budget; c) benchmarks measuring training and inference efficiency.  & Technical \\ \cline{2-3}

~ & Consider measurable key performance indicators (KPIs) on reporting to SDGs for reduction of energy bills, increased RES integration and reduction of environmental footprint. & Technical \\ \cline{2-3}

~ & Consider the AI Project Canvas (more general) or Human-Centered AI Canvas \cite{MailletAlberic2019IntroducingCanvas} (focused on human / social wellbeing) & Non-technical \\ \cline{2-3}

~ & Revisit and comply with international \cite{IEA2024IEAAgency, IRENA2024IRENAAgency} and EU directives \cite{Erbach2015EnergyStrategy} regarding RES \cite{EuropeanParliament2018Directiverecast}, energy efficiency \cite{EuropeanParliament2010DirectiveBuildings} and climate change \cite{EuropeanParliament2018RegulationAction} & Non-technical \\ 

\bottomrule
\end{tabular*}
\caption{Risks and proposed methods with respect to societal and environmental well-being}
\label{tab:wellbeing}
\end{table}

\subsubsection{Accountability}
Accountability refers to procedures relating to the responsibility during the development, deployment and use of AI systems. It involves risk management, internal and external auditing capabilities of the AI system and the management of trade-offs. This requirement can be considered as a superset of the rest of requirements and therefore can be easily managed if put into practice together with the previous ones. Table \ref{tab:accountability} indicatively lists risks and proposed methods relating to accountability.

\begin{table}[htbp]
\centering
\renewcommand{\arraystretch}{1.3} 
\begin{tabular*}{\linewidth}{p{1.4cm}|p{4.6cm}|p{1.6cm}}
\toprule
\textbf{Risks} & \textbf{Methods \& Tools} & \textbf{Method Type} \\ \hline
\midrule

\multirow{1}{\linewidth}{Allegation, Prosecution, Opaque development processes, Distrust
} 
 & Conduct algorithm impact assessments & Non-technical \\ \cline{2-3}

~ & Conduct internal and external audits & Non-technical \\ \cline{2-3}

~ & Establish a TAI review board  & Non-technical \\ \cline{2-3}

~ & Strong documentation regarding trade-offs (e.g. accuracy/explainability, privacy/safety) & Technical \\ 

\bottomrule
\end{tabular*}
\caption{Risks and proposed methods with respect to accountability}
\label{tab:accountability}
\end{table}

\subsection{Application methodology}
Aiming to draw a common approach regarding TAI services, in this section we propose a methodological procedure, named E-TAI, for the application of TAI in the development of AI systems in the energy sector. The methodology is based on the ALTAI, as well as on the proposed guidelines for the identification and management of ethical risks, as described in the previous section. 

After each assessment cycle of the AI system (their duration and number depends on the internal development and evaluation procedures of the stakeholders), the most important risks for each one of the TAI requirements (described in the previous section) should be identified and reported. Moreover, it is suggested that the stakeholders and developers of the AI system seek several technical and non-technical methods and tools that can help addressing the identified risks, alongside specific actions that are proposed to mitigate them. 

\begin{figure}[htbp]
\centering
\includegraphics[height=10.1cm, keepaspectratio]{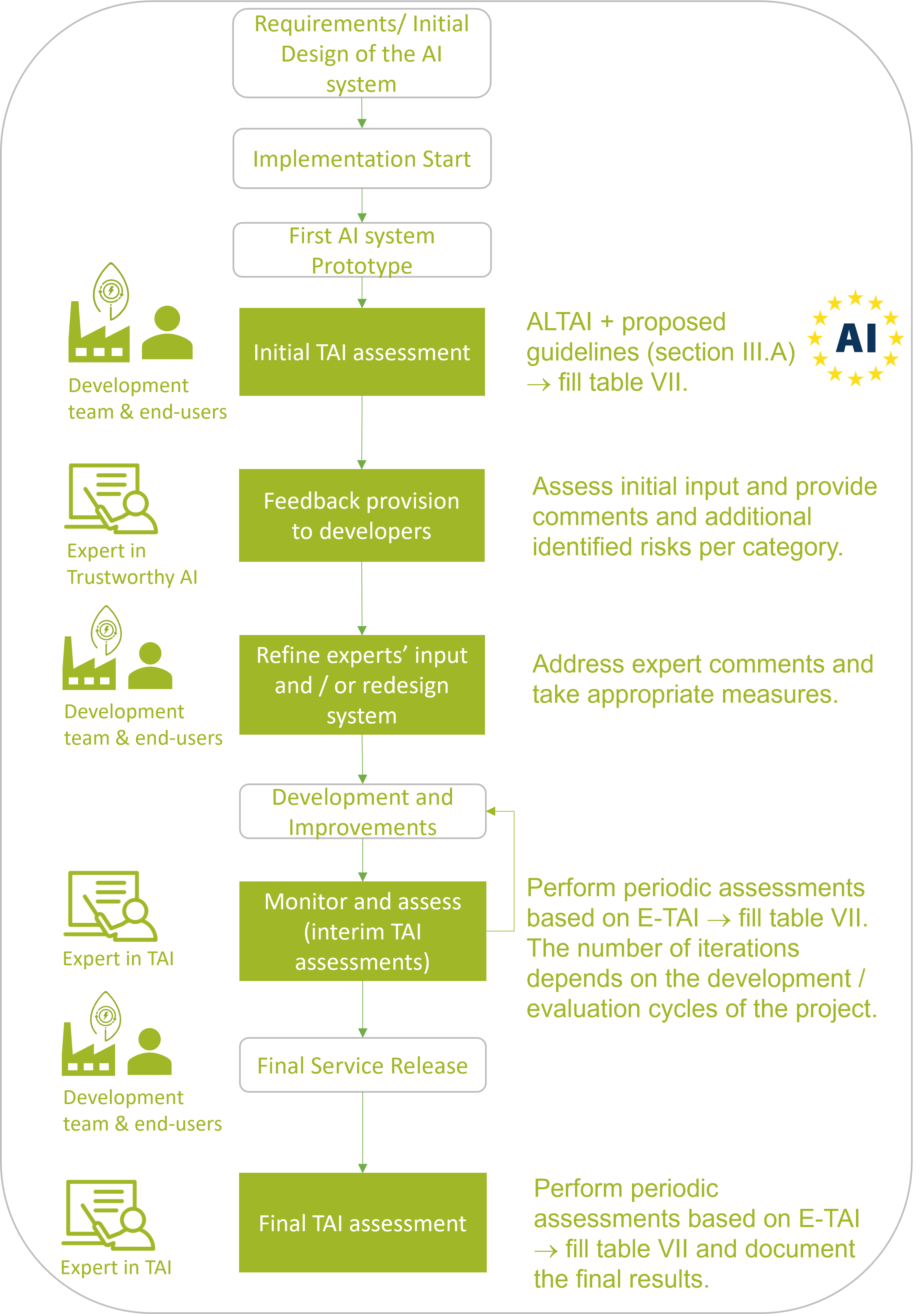}
\caption{Procedure for assessing an energy sector AI system (E-TAI)}
\label{fig:schema}
\end{figure}

The overall procedure is illustrated in Fig.~\ref{fig:schema} and is recommended for all stakeholders involved in the development and deployment of AI systems (end-users, developers) within the energy sector. Most importantly, the methodology also recommends the optional involvement of TAI experts to assist the evaluation of the AI system. During each iterative evaluation cycle, the stakeholders of the AI systems are highly encouraged to fill the ALTAI questionnaire for each requirement and optimize their development and future planning accordingly. 

\subsection{Evaluation}
E-TAI has been applied and evaluated in the I-NERGY EU project. Specifically, it was applied by 9 pilots in 15 different use cases to assist the development of 19 successful TAI services for the energy sector. The services were related to energy networks, distributed energy resources, and energy efficiency. The resources required for the application of E-TAI depend on the scope of the AI system, however, empirically, at least one developer (full time), one domain expert (supervision), and one TAI expert are recommended (occasional support). Evaluation is also underway in EnerShare and iTrust6G projects.

\section{Conclusions and future work}
In the present study we proposed E-TAI, a novel framework for evaluating TAI systems in the energy sector consisting of a set of supportive guidelines, alongside an iterative application methodology. Additionally, an analysis of the current landscape with respect to regulations and initiatives within the EU provided insights on the the status quo in TAI. The proposed framework is directed to EPES stakeholders involved in the utilization of AI systems within smart energy contexts and can be used as a handbook for the iterative assessment and evaluation of said systems with respect to trustworthiness.

With respect to future work, it is of utmost importance that the AI act is officially voted (pending EU Parliament's vote on March 2024) and that E-TAI is refined and updated accordingly to be in alignment with the new forthcoming regulation. Additionally, any other directives, regulations, and standards such as CEN-CENELEC and the ISO should be constantly monitored and further specialized to the energy domain, coming up with even more standardized approaches. Ultimately, E-TAI is also under modification and evaluation in the iTrust6G project, whose results are pending future publication, seeking its extension to new domains.

\section*{Acknowledgment}
This work has been funded by the European Union’s Horizon Europe research and innovation program under the iTrust6G project, grant agreement No. 101139198.

\printbibliography

\end{document}